\documentclass{mem}
\usepackage{natbib}
\usepackage{txfonts}
\usepackage{balance}
\usepackage{graphicx}
\begin{document}

\title{
Jets in high-mass microquasars
}

   \subtitle{}

\author{
M. Perucho\inst{1,2} 
          }

  \offprints{M. Perucho}
 
\institute{
Departament de Matem\`atiques per l'Economia i l'Empresa, Universitat de Val\`encia, 
Av. Tarongers, s/n, 46022, Val\`encia, Valencian Country, Spain.
\and
Departament d'Astronomia i Astrof\'{\i}sica, Universitat de Val\`encia, Av. Dr. Moliner, 50,
46100, Burjassot, Valencian Country, Spain.
\email{manel.perucho@uv.es}
}

\authorrunning{Perucho}

\titlerunning{Jets in high-mass microquasars}

\abstract{The morphologies of detected jets in X-ray binaries are almost as diverse as their number.
This is due to different jet properties and ambient media that these jets encounter. 
It is important to understand the physics of these objects and to obtain information about possible sites suitable for
particle acceleration in order to explain the observations at very high energies. Here I present the results obtained 
from the first relativistic hydrodynamical simulations of jets in high-mass microquasars. Our results allow us to make 
estimates for the emission originated in different sites of the whole structure generated by the jets. These works 
represent a first step in trying to obtain a deeper understanding of the physics and emission processes related with 
jets in high-mass microquasars.
}
\maketitle{}

\section{Introduction}

High-mass X-ray binaries (HMXB) are the parent population of high-mass microquasars (HMMQ), the latter being HMXB with jet
activity. Jets of X-ray binaries (microquasars) are produced close to the compact object (black hole or neutron star) via
ejection of material accreted from the stellar companion. Jets in HMMQs propagate in a medium filled by the stellar wind of
the massive primary star of the system. At the scales of the binary, the stellar wind can significantly affect the dynamics
of the jet, trigger shocks suitable for particle acceleration, and under some conditions the jet may be disrupted by a
combination of asymmetric recollimation shocks and hydrodynamical instabilities (see Perucho \& Bosch-Ramon 2008, Perucho et
al. 2010). On the other hand, powerful jets can propagate farther, and once at distances from the injection point much bigger
than the separation distance ($z\gg d_{\rm orb}$, where $z$ is the axial coordinate and distance to the compact object).
Depending on the age of the HMXB, the jet will go through the Supernova remnant (SNR) created by its progenitor or, at later
stages, the shocked stellar wind, and then the ISM  (Bosch-Ramon et al. 2011). After crossing these regions, the jet can
propagate as far and fast as the inertia of the swept ISM permits, and eventually inflates a wide
cocoon that pushes a shell of shocked ISM (Bordas et al. 2009). The occurrence of collisionless shocks in microquasar jets
can lead to efficient particle acceleration (e.g.,
Rieger et al. 2007, Perucho \& Bosch-Ramon 2008, Bordas et al. 2009, Bosch-Ramon
et al. 2011) and non-thermal emission of synchrotron and inverse Compton origin and, possibly, from proton-proton collisions
(see Bosch-Ramon \& Khangulyan 2009 and references therein). 

To understand the dynamics of these jets, numerical simulations of their propagation were performed (Peter \& Eichler
1995; Vel\'azquez \& Raga 2000; Perucho \& Bosch-Ramon 2008, Bordas et al. 2009, Perucho et al. 2010, Bosch-Ramon et al.
2011) together with several analytical treatments (e.g., Heinz \& Sunyaev 2002; Kaiser et al. 2004; Araudo et al. 2009). In
this contribution, I review the results from our simulations of microquasar jets at different scales.

\section{A physical picture obtained from the simulations}

For our study, we used a finite-difference code named \textit{Ratpenat}, which solves the equations of relativistic hydrodynamics in conservation form using high-resolution-shock-capturing methods (see Perucho et al.~2010). 

\subsection{Binary region}

\begin{figure*}[]
 \resizebox{\hsize}{!}{\includegraphics[clip=true]{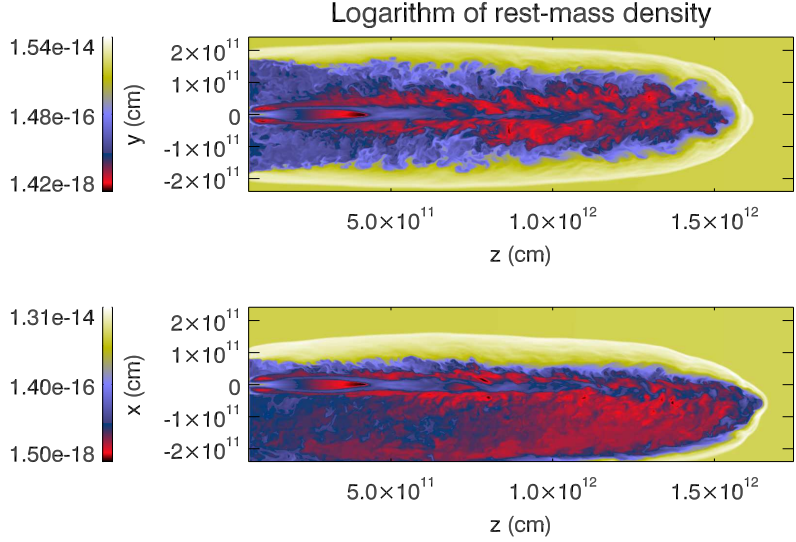}}
 \caption{\footnotesize Cuts of logarithm of rest mass density in the plane defined by the binary star and the jet (top) and perpendicular to it (bottom) for a jet with power $L_j = 10^{36}$~erg/s after 1000 s.
 }
 \label{fig1}
 \end{figure*}

  The jet is injected supersonically (see Perucho \& Bosch-Ramon 2008, Perucho et al. 2010) in the ambient and generates a bow-shock. It reproduces the traditional morphology of supersonic outflows, i.e., the jet flow reaches the terminal (reverse) shock at the head of the jet gas and fills a cavity of shocked jet material that surrounds the beam. The cocoon is surrounded by the shocked ambient medium. Both jet and ambient shocked material mix in the contact discontinuity that separates the cocoon and the shocked ambient gas. However, inside the binary region ($z<3\times10^{12}$~cm), the ambient consists on a stellar wind from the companion massive star. In the case of massive stars (types OB), this wind can be powerful, with mass losses on the order of $10^{-6}$~$M_\odot$~yr${}^{-1}$ and typical velocities on the order of $2\times 10^8$~cm~s${}^{-1}$ (see Perucho \& Bosch-Ramon 2008 and references therein). In this situation, the bow-shock is impacted by this wind on the side where the companion star is with respect to the compact object at the moment of injection. On the other side, the bow-shock expands in the rarefaction generated by the wind in the same direction. This asymmetry causes pressure differences on both sides of the jet. As long as the jet is powerful ($L_j \geq 10^{37}$~erg/s), and the pressure jump in the shock is large enough, the asymmetry induced in the bow-shock shape is not significant. Figure~\ref{fig1} shows the last snapshot of the simulation for the case of a jet with power $L_j \geq 10^{36}$~erg/s. This image shows the influence of the wind on the shape of the bow-shock for weak jets and the jet helical morphology arising after the recollimation shock.

  The jet flow can be overpressured with respect to the surrounding cocoon and undergo an adiabatic expansion until it gets underpressured with respect to the ambient, recollimating in a strong reconfinement shock. Up to this shock, the large overpressure of the jet makes it insensitive to the asymmetry in pressure generated by the wind. After the reconfinement, the flow is decelerated and the aforementioned asymmetry generates helical Kelvin-Helmholtz instabilities
that affect the whole body of the jet. These instabilities may cause the disruption of jets with power $L_j \leq 10^{37}$~erg/s within the binary region. Disruption by instabilities causes mass entrainment, deceleration and an increase of the jet cross-section. 

\begin{figure*}[]
 \resizebox{\hsize}{!}{\includegraphics[clip=true]{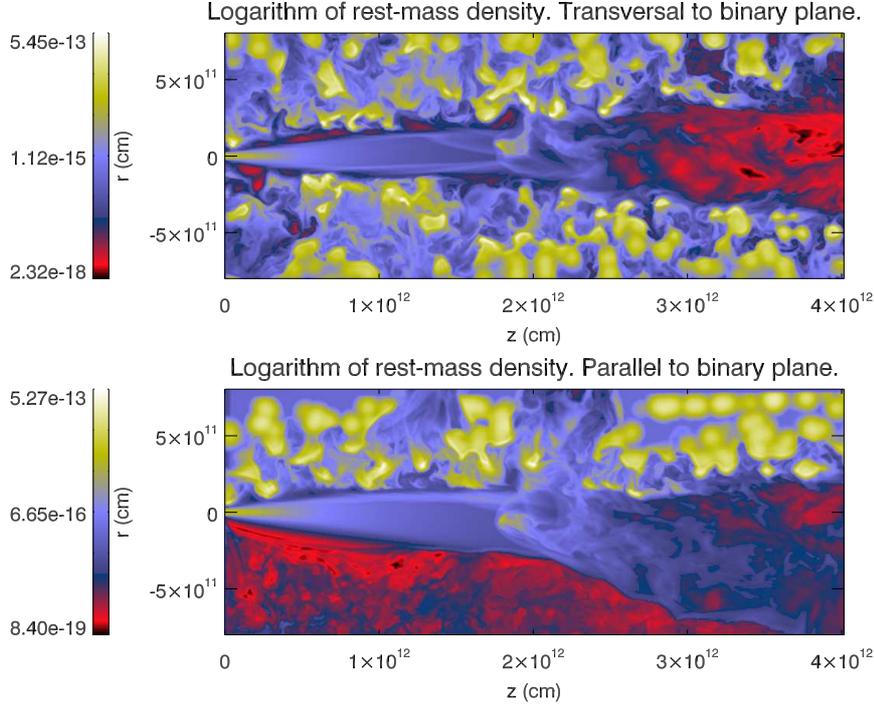}}
 \caption{ \footnotesize Cuts of logarithm of rest mass density in the plane defined by the binary star and the jet (top) and perpendicular to it (bottom) for a jet with power $L_j = 10^{37}$~erg/s. 
 }
 \label{fig2}
 \end{figure*}

The winds from massive stars are known to be clumpy and this feature could be of importance 
regarding the jet propagation, due to strong mass entrainment by dense clumps that propagate into the jet. We have performed simulations of an overpressured jets in direct contact with the wind (i.e., we implicitly assume that the bow shock is far away and the cocoon/backflow has been evacuated by the wind). In these simulations we have included thermal radiative cooling terms. The jet powers are $L_j = 3\times 10^{36}$~erg/s and $L_j = 10^{37}$~erg/s. Our results show strong entrainment and interaction with individual knots (see Fig.~\ref{fig2}). This, together with the wind thrust, causes a deviation of the jet's original direction of motion and efficient deceleration of the jet material.

\subsection{Outside the binary region}

  Once the jet is outside the binary region, at $z\sim3\times10^{18}$~cm~$\simeq 1$~pc, the ambient that it will 
encounter will depend on the age of the HMXB (Bosch-Ramon et al.~2011). In the case of a young HMXB, with an age of 
$t_{\rm src}\sim 3\times 10^4$~yr, the jet would encounter the stellar wind shocked by the shocked SNR, which to its 
turn drives a strong shock in the ISM. In this case, a jet with power $L_j=3\times10^{36}$~erg/s is strongly decelerated 
by the shocked ISM and it would tipically need several thousands of years to cross this region and reach the ISM.
\begin{figure}[]
\resizebox{\hsize}{!}{\includegraphics[clip=true]{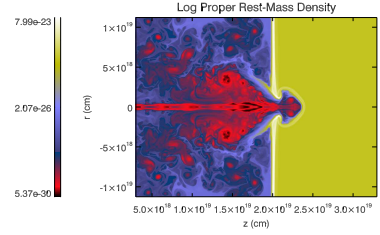}}
 \caption{\footnotesize Rest mass density maps showing the interaction of a jet with 
power $L_j = 3\times 10^{36}$~erg/s with the shocked stellar wind and shocked ISM.
 }
 \label{fig3}
 \end{figure}
  In the case of a HMXB with an age around $t_{\rm src}\sim 10^5$~yr, after crossing the shocked wind region, the jet is decelerated at the thin layer of shocked ISM. However, after a few thousand years, the jet head drills this layer completely and enters the ISM region (Fig.~\ref{fig3}).

  For an older source, with an age $t_{\rm src}\sim 10^6$~yr, the ISM shock velocity could become slower than the proper velocity of the system through the ISM ($v_p\sim 10^7$~cm/s). In this case, a bow-shock structure would form in the direction of $v_p$ and the shocked wind would be evacuated in the opposite direction with a similar velocity to its unshocked wind one $v_w \simeq 10^8$~cm/s. In this case, our simulations show that a jet with power $L_j=3\times10^{36}$~erg/s would be disrupted by the effect of the side impact of the shocked wind. As a result of this interaction, the jet is deviated and decelerated to the extent that it is not able to cross the shocked ISM, which acts as a wall for the jet material. 

\subsection{At last, the ISM}
  
Once the jet escapes all the previously mentioned interaction regions, it enters the ISM, where its evolution can be approximated as a jet propagating in a homogeneous ambient medium (Bordas et al. 2009). In such standard scenario, an overpressured jet generates a bow-shock in the ambient, and undergoes ones ore several reconfinement shocks due to its pressure difference with the surrounding cocoon. 

Bordas et al. (2009) modelled this evolution and studied the emission from this kind of systems with ages between $10^4$ and $10^5$~yrs. The calculations of the emission was based on the previous description of the system, which relies on both analytical and numerical work.

\section{Particle acceleration sites and emission}

In the previous sections, we have described some important processes in the evolution of jets in HMXBs. In all of them,
strong interactions take place, which are suitable locations for particle acceleration: a) Bow-shock, reverse  and
reconfinement shock in the binary region and in the ISM, and b) strong interaction of the jet with the shocked SNR and/or ISM at different ages of the HMXB.

Our results put limits to the possible detection of emission from jets in HMXBs in the different studied spatial and temporal
scales. The estimates we make for the different situations show that the radiation produced at recollimation shocks inside
the binary region, $\sim 10^{12}$~cm, could explain the observed high-energy emission from several high-mass X-ray binaries (HMXB) (see, e.g., Bosch-Ramon \& Khangulyan 2009 and references therein). 

Once the jet is far outside the binary region, and depending on its age, it could be embedded in the hot shocked wind/SNR
ejecta. At this stage, significant non-thermal emission can be generated in the reverse shock and cocoon region, mainly
radio and X-rays, and also very high-energy photons, via synchrotron radiation and inverse compton scattering of
photons from the companion star. On the contrary, as the jet propagates through the hot ambient, there is no forward
bow-like shock until it reaches the shocked ISM. In that stage, the jet is pressure confined inside the hot
shocked wind and/or SNR and this leads to pinching and reconfinement. Thus, particle acceleration may take place in
these regions. See the discussion in Bosch-Ramon et al. (2011) about the implications of these results for sources like
SS433 and Cygnus X-1.

In the case of jets propagating in the ISM, Bordas et al. (2009) showed that MQs could be extended non-thermal emitters from radio to gamma-rays, with emission being generated in the shocked ambient medium, reverse shock and in reconfinement shock(s) along the jet. Although these results strongly depend on the non-thermal luminosity fraction and on the magnetic field assumed for the calculations, the model is able to account for sources from which radiation could be detected and those from which it would not, mainly depending on jet power, age and ambient density. In general, the emitted flux is larger for larger jet power and ambient density, whereas the changes due to the source age are frequency and radiative-process dependent.

\begin{acknowledgements}
I would like to thank Valent\'{\i} Bosch-Ramon for the long discussions that have resulted in my present knowledge about microquasars. I acknowledge support by the Spanish ``Ministerio de Ciencia e Innovaci\'on'' (MICINN) grants AYA2010-21322-C03-01, AYA2010-21097-C03-01 and CONSOLIDER2007-00050, and by the ``Generalitat Valenciana'' grant  ``PROMETEO-2009-103''.    
\end{acknowledgements}

\bibliographystyle{aa}

%\bigskip
%\bigskip
%\noindent {\bf DISCUSSION}
%
%\bigskip
%\noindent {\bf JIM BEALL:} What is the velocity of the jets with respect to the various kinetic luminosities?
%
%\bigskip
%\noindent {\bf MANEL PERUCHO:} The velocities of the jets we simulate are all in the range between 0.3 and 0.6 c. 
%
%\bigskip
%\noindent {\bf SERGIO COLAFRANCESCO:} Is your code able to catch evidence of tearing instability in the jet plasma?
%
%\bigskip
%\noindent {\bf MANEL PERUCHO:} The code we have been using to perform these simulations (Ratpenat) is a RHD code. We are thus able to catch all macroscopic hydrodynamic instabilities, but not magnetic ones. 

\end{document}